\documentclass[aps,12pt,tightenlines,nofootinbib,showpacs]{revtex4}

\usepackage{graphicx,amsmath}

\newcommand{\textav}[1]{\langle#1\rangle}
\newcommand{\av}[1]{\left\langle#1\right\rangle}
\newcommand{\cum}[1]{\langle\!\langle#1\rangle\!\rangle}
\newcommand{\rcum}[1]{\Bigl\langle\!\!\Bigl\langle#1\Bigr\rangle\!\!\Bigr\rangle_r}
\newcommand{\sfrac}[2]{\mbox{$\frac{#1}{#2}$}}
\newcommand{\dd}[3][]{\frac{\partial^{#1}#2}{\partial#3^{#1}}}

\begin{document}

\title{Gaussian approximation to single particle correlations at and
below the picosecond scale for Lennard-Jones and nanoparticle~fluids}

\author{R. van Zon$^{1,3}$, S. S. Ashwin$^{2,3}$ and E. G. D. Cohen$^3$}

\affiliation{$^1$ Chemical Physics Theory Group, Department of Chemistry,
  University of Toronto, 80 St.\ George Street, Toronto, Ontario M5S
  3H6, Canada}

\affiliation{$^2$ Department of Chemistry, University of Saskatchewan, 110
  Science Place, Saskatoon, Saskatchewan S7N 5C9, Canada}

\affiliation{$^3$ The Rockefeller University, 1230 York Avenue, New York,
  New York, 10065-6399, USA}

\date{28 March 2008}

\begin{abstract}
To describe short-time (picosecond) and small-scale (nanometre)
transport in fluids, a Green's function approach was recently
developed. This approach relies on an expansion of the distribution of
single particle displacements around a Gaussian function, yielding an
infinite series of correction terms.  Applying a recent theorem [Van
Zon and Cohen, J.\ Stat.\ Phys.\ {\bf 123}, 1--37 (2006)] shows that
for sufficiently small times the terms in this series become
successively smaller, so that truncating the series near or at the
Gaussian level might provide a good approximation. In the present
paper, we derive a theoretical estimate for the time scale at which
truncating the series at or near the Gaussian level could be supposed
to be accurate for equilibrium nanoscale systems.  In order to
numerically estimate this time scale, the coefficients for the first
few terms in the series are determined in computer simulations for a
Lennard-Jones fluid, an isotopic Lennard-Jones mixture and a
suspension of a Lennard-Jones-based model of nanoparticles in a
Lennard-Jones fluid.  The results suggest that for Lennard-Jones
fluids an expansion around a Gaussian is accurate at time scales up to
a picosecond, while for nanoparticles in suspension (a nanofluid), the
characteristic time scale up to which the Gaussian is accurate becomes
of the order of five to ten picoseconds.
\end{abstract}


\maketitle

\section{Introduction}

\label{introduction}

Small clusters of particles suspended in a fluid occur in many forms,
from nanoparticles \cite{Choietal02, Hwangetal04,TangAdvani06},
quantum dots\cite{Chenetal05} and colloidal suspensions
\cite{Verbergetal97} to biomolecules such as globular proteins
\cite{TenWoldeFrenkel97, Pellicaneetal04}. Such nanoclusters have a
variety of applications, from material coatings to drug delivery by
hollow clusters.  Both the individual behaviour of nanosized
particles\cite{RothBalasubramanya00,BalettoFerrando05, Szwackietal07}
as well as their collective behaviour, such as the increased heat
conductance in dilute suspensions of nanoparticles (so-called
nanofluids)\cite{Choietal02}, have received considerable
attention\cite{BarratHansen}.

For the purpose of studying small length scale and short time
classical transport phenomena which occur in nanosystems, a Green's
function approach was introduced by Kincaid\cite{kincaid}. This
approach has the promise of being able, in principle, to describe
transport phenomena on all time and length scales, unlike
hydrodynamics. The main idea of the theory is to describe the
evolution of fluid properties such as its energy, momentum and number
density in terms of Green's functions. The application of these
Green's functions to nanosystems and systems where time scales at
picoseconds or less are important, has been an area of some
interest\cite{kincHydro,kincNano,kincHeat, VanzonCohenarchive}.  In
these cases, the Green's functions were expanded around a Gaussian
distribution plus an infinite series of corrections, a finite
truncation of which yielded excellent agreement with simulations. Even
just the Gaussian itself was found to be a reasonable approximation to
the Green's functions. An explanation for this could be that the
series of corrections has fast convergence, but at that point, it was
not known why that this could be the case. Since the Gaussian
description is much simpler than the full Green's function, one would
like to know when fast convergence occurs and when taking the Gaussian
approximation suffices.  A preliminary answer to this question was
found in Ref.\,\onlinecite{BrusselsPaper}, namely, that for the motion
of a single particle in an equilibrium pure Lennard-Jones (LJ) fluid,
the Gaussian approximation can be used up to time scales of the order
of a picosecond.

One of the applications of the Green's function approach is mass
transport in liquids and liquid mixtures. For that case, the Green's
functions are essentially the probability distribution functions of
displacements (in a time $t$) of single particles of the different
components \cite{VanzonCohenarchive}.  Thus it is not too surprising
that the Green's functions can be expressed in terms of the cumulants
of this distribution. These cumulants measure the correlations of the
displacement of a single particle, in particular, they measure the
departure of the correlations from Gaussian behaviour. As will be
discussed in more detail below, a recent theorem regarding these
cumulants implies that when the Green's functions are expanded around
a Gaussian distribution, the correction terms to the Gaussian term are
proportional to increasing powers of $t$ for short (initial) times
$t$\cite{VanZonCohen05}.  Analytic expressions for the coefficients in
front of the powers of $t$ were also derived in
Ref.\,\onlinecite{VanZonCohen05}.  The values of the first two
numerical coefficients are here of particular interest, because they
can be computed numerically and, as show in Sec.\,\ref{sec:eq} can
then be used to find estimates of the physical time scales below which
the expansion of the Green's function around the Gaussian term yields
useful results, as appeared to be the case in
Refs.\,\onlinecite{kincaid,kincHydro,kincNano,kincHeat}. Numerical values for these
coefficients will be presented in this paper for various equilibrium
LJ-based systems, including nanoparticles in a suspension of LJ
particles.  We will present the resulting orders of magnitude of the
relevant time scales on which the first few terms in the series
decrease.  Non-equilibrium systems will be studied in future work.

\section{Systems}

\label{systems}

Three systems were studied, namely a pure LJ fluid, an isotopic binary
mixture of LJ particles (in which context the study of short time
displacements arose\cite{VanzonCohenarchive}), and a suspension of
nanoparticles in a LJ fluid.

In the isotopic binary LJ mixtures, there are $N_A$ particles of mass
$m_A$ and $N_B$ particles of mass $m_B$ in a box of size $L^3$, such
that the number density is $\rho=(N_A+N_B)/L^3$. For the pure LJ
fluid, one sets $N_B=0$.  The positions and velocities of the
particles will be denoted by $\mathbf r_{\lambda i}$ and $\mathbf
v_{\lambda i}$, respectively, where $\lambda=A$ or $B$ and $i$ is a
particle index, which runs from 1 to $N_A$ if $\lambda=A$ and from 1
to $N_B$ if $\lambda=B$.  By definition, in an isotopic mixture all
pair interaction potentials are the same for all components, but their
masses are different.  The inter-atomic potential between the
particles is the LJ potential
\begin{equation}
  V_{AA}(r) = V_{AB}(r) = V_{BB}(r) 
  = 4\epsilon\left[
              \left(\frac{\sigma}{r}\right)^{12}
             -\left(\frac{\sigma}{r}\right)^6
             \right],
\label{VLJ}
\end{equation}
where $r$ is the distance between two particles, and $\sigma$ and
$\epsilon$ are the same for all pairs of particles.

All quantities reported are in LJ units: length in units of $\sigma$,
temperature in units of $\epsilon/k_B$, density ($\rho$) in units of
$\sigma^{-3}$ and time in units of $\tau_{LJ}=(\sigma^2
m_A/\epsilon)^{1/2}$, where $m_A$ is the mass of an $A$-particle.  In
other words, we will use units in which $\sigma=1$, $k_B=1$,
$\epsilon=1$, and $m_A=1$.  Although these are arbitrary units, to
understand the physical consequences of our results, we use the LJ
parameters of Argon as a reference. In that case, one unit of time
corresponds to $\tau_{LJ}=2.16\times10^{-12}$ seconds, while one unit
of length corresponds to $\sigma=0.34$\,nm\cite{frenkel,rapaport}.

As mentioned above, apart from the pure LJ fluid and the isotopic
binary LJ fluid mixture, a third system which will be studied, namely,
a suspension of nanosized particles in a fluid, often called a
nanofluid.  One can obtain this system from the binary isotopic LJ
fluid mixture by changing the $B$ particles to much larger, nanosized
particles while the $A$ particles remain regular LJ particles, and
changing the potentials $V_{AB}$ and $V_{BB}$ in the following
way. Each nanoparticle is represented as a spherical cluster of radius
$R$ with a smoothed uniform distribution of $M$ LJ particles as
proposed in Refs.\,\onlinecite{RothBalasubramanya00} and
\onlinecite{VanZonunpublished}.  Since we are only after typical time
scales for which the expansion presented in Sec.\,\ref{expand} below
is valid, we restrict ourselves here to this simple nanoparticle
model. For simplicity, we therefore take the strength of the LJ
potential between the constituent LJ particles of the nanoparticles
and the fluid particles to be the same, and the mass of the
constituent LJ particles of the nanoparticle is also taken to be equal
to that of the fluid particles. $R$ will range from $1$ to $6$ in LJ
units, i.e.\ from 0.34 nm to 2 nm (which is a typical size of a
quantum dot\cite{Chenetal05}), while $M$ will be chosen such that for
$R=0$, the nanoparticle reduces to a single LJ particle ($M=1$) while
for large $R$ the density of LJ particles within the nanoparticle
approaches one. This can be accomplished by choosing $M$ to be
$1+R^3$, leading to a maximum mass ratio of 217 between the
nanoparticles and the fluid LJ particles.  One can show that the
result of integrating the LJ potentials corresponding to all the
points in the spherical nanoparticle is that a nanoparticle interacts
with a fluid LJ particle through the
potential\cite{RothBalasubramanya00, VanZonunpublished}
\begin{equation}
  V_{AB}(r) = 4M\left[
    \frac{\frac43R^6+\frac{36}{5}R^2r^4}{(r^2-R^2)^9}
  + \frac{1}{(r^2-R^2)^6}
  - \frac{1}{(r^2-R^2)^3},
\right]
\label{VAB}
\end{equation}
where $r$ is the distance between the centre of the nano particle and
the LJ fluid particle, while the interaction potential between two
nanoparticles is given by\cite{VanZonunpublished}
\begin{align}
V_{BB}(r) &= 4M^2 \Bigg[
  \frac{r^{10}-\frac{8}{5}R^2r^8+\frac{216}{25}R^4r^6
  -\frac{1504}{75}R^6r^4+\frac{13696}{525}R^8r^2
  -\frac{512}{35}R^{10}}{r^8(r^2-4R^2)^7} 
\nonumber\\*&\qquad\qquad
  -\frac{3}{8R^4}\left\{
                   \frac{r^2-2R^2}{r^2(r^2-4R^2)}
                  -\frac{1}{4R^2}\ln\left(1-\frac{4R^2}{r^2}\right)
                 \right\}
  \Bigg],
\label{VBB}
\end{align}
where $r$ is the distance between the centres of the nanoparticles.
Note that because of the much larger size of the nanoparticles, far
fewer will fit into a system of given volume than $B$ particles fit in
an isotopic LJ mixture of only LJ particles.

The systems studied in this paper are all in canonical equilibrium,
i.e., their distribution function $\rho_{eq}({\mathbf \Gamma})$ in
phase space (${\mathbf \Gamma}=\{\mathbf r_{\lambda i}$,$\mathbf
v_{\lambda i}\}$) is given by:
\begin{equation}
  \rho_{eq}(\mathbf\Gamma) = e^{-H(\mathbf\Gamma)/T}/Z,
\label{rhoeq}
\end{equation}
where $Z=\int \exp[-H({\mathbf\Gamma})/T]{\rm d}{\mathbf \Gamma}$ is
the partition function, $T$ is the temperature, and $H$ is the
Hamiltonian which is of the form
\begin{equation}
  H(\mathbf\Gamma) =
 \sum_{\lambda=A,B}\sum_{j=1}^{N_\lambda}\frac{m_\lambda|{\mathbf v_{\lambda
 j}}|^2}{2} + U,
\label{Ham}
\end{equation}
where $U$ is a sum of pair potentials:
\begin{equation}
U = \sum_{\lambda=A,B}\sum_{i=1}^{N_\lambda}
  \sum_{\mu=A,B}\sum_{j=1}^{N_\mu}{\Big.}' \sfrac12
  V_{\lambda\mu}(|\mathbf r_{\lambda i}-\mathbf r_{\mu j}|),
\label{U}
\end{equation}
where the prime excludes equal particles (i.e., $\lambda=\mu$ and
$i=j$) and the $V_{\lambda\mu}$ are of the form given in
Eqs.\,\eqref{VLJ}--\eqref{VBB} above. Finally, we remark that the
equations of motions are given by
\begin{equation}
{\bf \dot{r}}_{\lambda i} = \mathbf v_{\lambda i};\qquad
{\bf \dot{v}}_{\lambda i} = -\frac{1}{m_\lambda}\dd{U}{\bf r_{\lambda i}}.
\label{eq9}
\end{equation}

\section{Green's functions and cumulants}

\label{expand}

We will now briefly review the Green's functions approach and its
connection with the distribution of single particle displacements.
For mass transport processes, the number density $n_\lambda({\bf r},
t)$ of a specific component $\lambda$ at position ${\bf r}$ at time
$t$ can be written as
\begin{equation}
  n_\lambda({\bf r},t) 
  = \int\!{\rm d}{\bf r}'\: G_\lambda({\bf r}-{\bf r}',{\bf r}',t)
  n_\lambda({\bf r}',0),
\end{equation}
where $G_\lambda({\bf r},{\bf r}',t)$ is the Green's function for
component $\lambda$ ($A$ or $B$ for a binary mixture), which is
defined as\cite{kincaid,VanzonCohenarchive}
\begin{equation}
  G_\lambda({\bf r},{\bf r}',t) 
  = \frac{
    \textav{\delta[{\bf r}'+{\bf r}-{\bf r}_{\lambda i}(t)]
      \delta[{\bf r}'-{\bf r}_{\lambda i}(0)]}_{\mathrm{is}}
  }{
    \textav{\delta[{\bf r}'-{\bf r}_{\lambda i}(0)]}_{\mathrm{is}}
  }
  ,
\label{Gdef}
\end{equation}
where ${\bf r}_{\lambda i}(t)$ is the position of the $i$th particle
of component $\lambda$ at time $t$ and the average
$\textav{}_{\mathrm{is}}$ is over a (possibly non-equilibrium)
initial state (``is''), which has to be specified for the particular problem
that one wants to study. The Green's function $G_\lambda({\bf r},{\bf
r}',t)$ can be interpreted as the probability that particle $i$ of
component $\lambda$ was displaced over ${\bf r}$ in a time $t$ given
that it started at ${\bf r'}$. Note that the Green's functions do not
depend on $i$ because particles of the same kind are
indistinguishable.

Although the Green's function approach is aimed primarily at
non-equilibrium systems, we will restrict ourselves here only to
equilibrium systems, because the time scales for the validity of the
expansion to be presented below are expected to be similar in
equilibrium and not-too-far-from-equilibrium systems, and the
equilibrium system is much easier to deal with from a numerical point
of view. In the equilibrium case, the Green's functions become
independent of ${\bf r'}$ because the system is homogeneous and are
then identical to the Van Hove self-correlation functions
$G^\lambda_s({\bf r},t)$ (with $\lambda$ a component) defined
as\cite{hansen}
\begin{equation}
  G^\lambda_s({\bf r },t) 
  =  \frac{1}{N_\lambda}
     \sum_{i=1}^{N_\lambda}
     \left\langle 
     \delta[{\bf r} + {\bf r}_{\lambda i}(0)-{\bf r}_{\lambda i}(t)]
     \right\rangle,
\label{Gsdef}
\end{equation} 
where the subscript $s$ refers to $G_s^\lambda$ being a
self-correlation function of a single particle.  The average
$\langle\,\rangle$ is here taken over the canonical equilibrium
ensemble $\rho_{eq}$ given in Eq.\,\eqref{rhoeq}.  To see that Eq.\,\eqref{Gsdef}
is the equilibrium variant of Eq.\,\eqref{Gdef}, note that each term on the
right-hand side of Eq.\,\eqref{Gsdef} gives the same contribution to the sum
due to the indistinguishability of particles of the same
component. Thus one can also write
\begin{equation}
  G^\lambda_s({\bf r },t) 
  =  \left\langle 
     \delta[{\bf r} + {\bf r}_{\lambda 1}(0)-{\bf r}_{\lambda 1}(t)]
     \right\rangle,
\label{Gs}
\end{equation} 
where particle $1$ of component $\lambda$ is used as a representative
particle of that component.  The expression for the Van Hove
self-correlation function in Eq.\,\eqref{Gs} coincides with that for the
Green's function in Eq.\,\eqref{Gdef} in cases where the Green's functions
have no ${\bf r}'$ dependence, i.e., in equilibrium. Note that like
the Green's function, the Van Hove self-correlation function
$G^\lambda_s({\bf r },t)$ can therefore be interpreted as the
probability that a single fluid particle of component $\lambda$ has
experienced a displacement {\bf r} in a time $t$.

The Fourier transform of the Van Hove self-correlation function is the
self-scattering function $F^\lambda_s({\bf k},t)$ \cite{hansen}, which
is given by:
\begin{equation}
  F^\lambda_s(k,t) 
  = \left\langle e^{ik{\bf k}\cdot[{\bf r}_{\lambda1}(t)-{\bf
        r}_{\lambda 1}(0)]}
    \right\rangle
  =\left\langle e^{ik\Delta x_{\lambda 1}(t)}\right\rangle. 
\end{equation} 
Here ${\bf k}=k{\bf \hat{k}}$ is a wavevector with length $k$ along the unit vector
$\hat{\mathbf{k}}$ and
\begin{equation}
\Delta x_{\lambda 1}(t)
= {\bf\hat k}\cdot[{\bf r}_{\lambda1}(t)-{\bf r}_{\lambda1}(0)]
\label{Deltaxdef}
\end{equation}
denotes the displacement of particle 1 of component $\lambda$ along
the direction $\hat{\mathbf{k}}$ at a time $t$.  The self-scattering
functions can be measured by incoherent neutron scattering
experiments\cite{neutronscattering}.

According to elementary probability theory \cite{kampen} one can
interpret $\log F^\lambda_s(k,t)$ as the cumulant generating function
of $\Delta x_{\lambda1}(t)$, where $\Delta x_{\lambda1}(t)$ is
considered to be a random variable, so that $F^\lambda_s(k,t)$ can be
written in the following form:
\begin{equation}
F^\lambda_s(k,t) = 
\exp{\left[\sum_{n=1}^{\infty} 
    \frac{\kappa^\lambda_{ n}}{n!}(ik)^{n}\right]}.
\label{eq3}
\end{equation}
Here $\kappa^\lambda_{ n}$ is called the $n$th cumulant of the
displacement $\Delta x_{\lambda1}(t)$.  The behaviour of these
cumulants as a function of time has been investigated in the context
of incoherent neutron scattering by Schofield \cite{schof} and Sears
\cite{sears}.  They showed that for equilibrium systems, the cumulants
($\kappa_n$ for $n=2$, 4, 6) have the following behaviour at small
times: $\kappa_2 \sim O(t^2)$, $\kappa_4 \sim O(t^8)$ and $\kappa_{6}
\sim O(t^{12})$, while the odd cumulants vanish in equilibrium. This
behaviour suggested a generalization, which has recently been obtained
for a certain class of physical systems as a Theorem
\cite{VanZonCohen05}. For a class of classical systems which includes
systems with smooth potentials\footnote{The LJ potential is not truly
smooth because it diverges at $r=0$. However, in equilibrium, this
point has a vanishingly small probability, so that the LJ potential
may be treated as effectively smooth.} in canonical equilibrium, it
was shown that the $\kappa^\lambda_n(t)$ have the following form:
\begin{equation}
\kappa^\lambda_{ n}=\left\{\begin{array}{ll}
  c^\lambda_nt^{n} + O(t^{n+1}) & \mbox{ for } n < 3  \\
  c^\lambda_nt^{2n} + O(t^{2n+1}) & \mbox{ for } n \geq 3.
	     \end{array}\right.
\label{eq4}
\end{equation}
where $c^\lambda_n$ are coefficients independent of $t$.  We see from
Eq.\,\eqref{eq3} that for sufficiently small wavevectors $k$,
$F^\lambda_s(k,t) \approx \exp[-\kappa^\lambda_{2} k^2/2]$. Since
$F^\lambda_s(k,t)$ is then approximately Gaussian in $k$, we would
expect that its inverse Fourier transform, the Van Hove
self-correlation function $G^\lambda_s(r,t)$, is also approximately
Gaussian in $r$. The corrections to the Gaussian behaviour of
$F^\lambda_s(k,t)$ are given by the terms in the series in Eq.\,\eqref{eq3}
with $n>2$. Taking the inverse Fourier transform of Eq.\,\eqref{eq3}, one
can show that the Van Hove self-correlation function is of the form of
a Gaussian plus corrections\cite{VanZonCohen05}:
\begin{equation}
G^\lambda_s(r,t)=\frac{\exp(-w^2)}{\sqrt{2\pi\kappa^\lambda_{2}}}\left[1 +
  \frac{\kappa^\lambda_{4}H_{4}(w)}{4!4[\kappa^\lambda_{2}]^2} +
  \frac{\kappa^\lambda_{6}H_{6}(w)}{6!8[\kappa^\lambda_{2}]^3}+\dots \right].
\label{eq5}
\end{equation}
Here $H_{n}$ is the $n$th Hermite polynomial, and
$w=r/\sqrt{2\kappa^\lambda_{2}}$ a dimensionless length.  Substituting
Eq.\,\eqref{eq4} in Eq.\,\eqref{eq5}, the Van Hove self-correlation function can
be expressed as a time series of the form:
\begin{equation}
G^\lambda_s(r,t)= \frac{\exp(-w^2)}{\sqrt{2\pi \kappa^\lambda_{2}}}
\left[1 +
  \frac{c^\lambda_4 m_\lambda^2 t^4}{96T^2} H_4(w)  
+ \frac{c^\lambda_6 m_\lambda^3 t^6}{5760T^3} H_6(w) +
  \dots\right]
\label{eq8}
\end{equation}
where we used that in equilibrium $c^\lambda_2 =
\textav{v_{\lambda1}^2} = T/m_\lambda$.

There are a few systems for which all the $c^\lambda_n$ for $n>2$ are
zero, leading to Gaussian Van Hove self-correlation functions. These
systems are the ideal gas and systems with only harmonic forces, whose
equations of motion are linear. For nonlinear systems, however, the
right-hand side of~Eq.\,\eqref{eq8} is a series in increasing even
powers of $t$. It is natural to expect that for a small enough $t$,
the successive terms in these series should rapidly decrease.  This
would mean that the series converges and that one could use a finite
number of terms, or even just the Gaussian, as a good approximation to
the whole series.  Applying the general rule that a series
$\sum_{n=0}^\infty a_n$ converges if
$\lim_{n\to\infty}|a_{n+1}/a_n|<1$ to the series in Eq.\,\eqref{eq8},
where $a_n\propto c^\lambda_{2n} t^{2n}$, it follows that the time
scale below which the decrease in the terms occurs depends critically
on the coefficients $c^\lambda_{2n}$, or in particular on ratios of
successive $c^\lambda_{2n}$ as $n$ approaches infinity. Infinitely
large values of $n$ are, of course, beyond the reach of numerical
computation but to get an estimate for the time scales, we numerically
evaluated $c_{2n}^\lambda$'s for the LJ liquid for finite $n$ up to
$n=3$ and the corresponding time scales for the decrease in the terms
of the series.

\section{Time scales}

\label{timescales}

As explained above, to numerically estimate the time scales up to
which the series expansion of the Van Hove self-correlation functions
$G^\lambda_s$ (with $\lambda=A$ or $B$) in Eq.\,\eqref{eq8} may converge or
at least be useful, we are interested in the first few terms of the
series. The terms in Eq.\,\eqref{eq8} which are of importance are then the
coefficients $c^\lambda_4$ and $c^\lambda_6$. Expressions for these
coefficients are derived in Sec.\,\ref{expr}, while in Sec.\,\ref{results} the
results of their numerical evaluation in simulations are presented.

For sufficiently small times $t$, every successive term in the series
in Eq.\,\eqref{eq8} would approach zero more rapidly than the previous term
because of a larger power of $t$ associated with it. This gives us a
simple relation to check when we could expect the terms in the series
to decrease.  The first estimate of a time scale, to be denoted by
$\tau^\lambda_G$, follows from the criterion that for
$t=\tau_G^\lambda$, the first term in the brackets in Eq.\,\eqref{eq8},
i.e.\ $1$, is of the same order of magnitude as the next term, i.e.\
${c^\lambda_4m_\lambda^2t^4}H_4(w)/(96T^2)$.  To find the order of
magnitude of the latter, we need an order of magnitude estimate for
$H_4(w)$, which we find as follows. The prefactor $e^{-w^2}$ in
Eq.\,\eqref{eq8} suggests that $w=O(1)$, since otherwise $G_s^\lambda$ would
be extremely small. The Hermite polynomial $H_4(w)$ contains no
physical parameters, only numerical factors which are also of $O(1)$,
so we conclude that $H_4(w)=O(1)$. The second term in Eq.\,\eqref{eq8} is
therefore of the order of the first term at $t=\tau_G$ with
${c^\lambda_4m_\lambda^2[\tau_G^\lambda]^4}/(96T^2)=\mathcal O(1)$,
yielding
\begin{equation}
\tau_G^\lambda 
= \left(\frac{96}{|c^\lambda_4|}\right)^{1/4} \sqrt{\frac{T}{m_\lambda}}
\label{tauG}
\end{equation}
This $\tau_G^\lambda$ expresses on what time scale a Gaussian
approximation to $G_s^\lambda$ will break down, while for time scales
somewhat less than to $\tau_G^\lambda$, the Gaussian distribution
could be supposed to be a good approximation.

The next simplest estimate of a time scale, to be denoted by
$\tau_*^\lambda$, is determined by the time $t=\tau_*^\lambda$ when
the second and third terms in the square brackets in Eq.\,\eqref{eq8} become
comparable, i.e., when:
\begin{equation}
\left|\frac{c^\lambda_4m_\lambda^2t^4}{96T^2} H_4(w)\right| =
\left|\frac{c^\lambda_6m_\lambda^3t^6}{5760T^3} H_6(w)\right|
\end{equation} 
which, using the same argument as above Eq.\,\eqref{tauG} to show that
typical values of $H_4(w)$ and $H_6(w)$ are $O(1)$, leads to
\begin{equation}
\tau_*^\lambda =
 \left(\frac{60|c^\lambda_4|}{|c^\lambda_6|}\right)^{1/2}
\sqrt{\frac{T}{m_\lambda}}.
\label{taustar}
\end{equation}
This $\tau_*^\lambda$ also defines a time scale below which the
subsequent terms in the series in Eq.\,\eqref{eq8} should decrease in
magnitude. Thus, for time scales sufficiently less than
$\tau_*^\lambda$, the $c_6^\lambda$ term can be neglected compared to
the $c_4^\lambda$ term in Eq.\,\eqref{eq8}, but for time scales larger than
$\tau_*^\lambda$, the $c_6^\lambda$ term certainly needs to be taken
into account.

One could in principle get additional time scale estimates
$\tau_n^\lambda$ by including higher order terms in Eq.\,\eqref{eq8} and
comparing the $n$th with the $n+1$st term. Note that then
$\tau_G^\lambda$ is equal to $\tau_1^\lambda$ and $\tau_*^\lambda$ is
equal to $\tau_2^\lambda$, respectively. If the limit $\tau^\lambda =
\lim_{n\to\infty}\tau_n^\lambda$ exists, the series in Eq.\,\eqref{eq8}
converges for all $t<\tau^\lambda$. In simulations, we cannot take
this limit, but we will see that $\tau_G^\lambda$ and $\tau_*^\lambda$
have similar orders of magnitude, suggesting that $\tau_G^\lambda$ and
$\tau_*^\lambda$ might be reasonable estimates of the actual time
scale of convergence of~Eq.\,\eqref{eq8}.

\section{Expressions for the coefficients $c^\lambda_4$ and $c^\lambda_6$}

\label{expr}

\subsection{General expressions}

We first discuss the analytical expressions for the coefficients
$c^\lambda_n$ in terms of the so-called multivariate cumulants based
on Ref.\,\onlinecite{VanZonCohen05}. The general relation between moments and
cumulants is given in \ref{A}. For short times, the
$\kappa^\lambda_{n}(t)$ have the form given by Eq.\,\eqref{eq4}, where for
$n \geq 3$ the scaling coefficients $c^\lambda_n$ are given
by\cite{VanZonCohen05}
\begin{equation}
c^\lambda_n = \mathop{
\underbrace{\sum_{n_1=0}^n\dots\!\!\sum_{n_{n+1}=0}^n}
_{\sum_{\gamma=1}^{n+1}n_\gamma=n} }_{\sum_{\gamma=1}^{n+1}\gamma
n_\gamma=2n} \frac{n!}{\prod_{\gamma=1}^{n+1}
[n_\gamma!(\gamma!)^{n_\gamma}]} \left\langle\!\! \left\langle
Y_{\lambda1}^{[n_{1}]};\dots;Y_{\lambda
n+1}^{[n_{n+1}]}\right\rangle\!\!\right\rangle .
\label{Clambdaj}
\end{equation}
Here, $\cum{Y_{\lambda1}^{[n_{1}]};\dots ;Y_{\lambda
n+1}^{[n_{n+1}]}}$ is a notation introduced in Ref.\,\onlinecite{VanZonCohen05}
for a multivariate cumulant, which is a multivariate moment with all
possible factorizations subtracted. In this notation, quantities
separated by semicolons are treated as separate random variables and
if a quantity has a superscript within square brackets, it denotes the
number of repetitions of that particular quantity, e.g.,
$\cum{Y^{[3]}_{\lambda1}} \equiv
\cum{Y_{\lambda1};Y_{\lambda1};Y_{\lambda1}}$ (see
\ref{A}). Furthermore, $Y_{\lambda\gamma}$ is defined as
\begin{equation}
   Y_{\lambda\gamma}
  = 
  \frac{d^\gamma\Delta x_{\lambda1}(t)}{dt^\gamma}\bigg|_{t=0},
\label{Ydef}
\end{equation}
with $\Delta x_{\lambda1}(t)$ defined in Eq.\,\eqref{Deltaxdef}.  Note
that we deviate here from the notation in
Ref.\,\onlinecite{VanZonCohen05}, where the cumulants were expressed
in terms of $X_{\lambda\gamma}=Y_{\lambda\gamma}/\gamma!$ instead of
in terms of $Y_{\lambda\gamma}$.

By writing out the sums in Eq.\,\eqref{Clambdaj} for $n=4$ and $n=6$, one
finds the following expressions for $c^\lambda_4$ and $c^\lambda_6$:
\begin{align}
c^\lambda_4 &=
\sfrac{1}{30}\cum{Y_{\lambda1}^{[3]};Y_{\lambda5}} +
\sfrac{1}{6}\cum{Y_{\lambda1}^{[2]};Y_{\lambda3}^{[2]}}
+\sfrac{1}{4}\cum{Y_{\lambda1}^{[2]};Y_{\lambda2};Y_{\lambda4}}
+\sfrac{1}{2}\cum{Y_{\lambda1};Y_{\lambda2}^{[2]};Y_{\lambda3}}
+\sfrac{1}{16}\cum{Y_{\lambda2}^{[4]}}
\label{c4gen}
\\
  c^\lambda_6 &=
  \sfrac{1}{840}\cum{Y_{\lambda1}^{[5]};Y_{\lambda7}}
  +\sfrac{1}{48}\cum{Y_{\lambda1}^{[4]};Y_{\lambda2};Y_{\lambda6}}
  +\sfrac{1}{24}\cum{Y_{\lambda1}^{[4]};Y_{\lambda3};Y_{\lambda5}}
  +\sfrac{5}{192}\cum{Y_{\lambda1}^{[4]};Y_{\lambda4}^{[2]}}
\nonumber\\*&\quad
  +\sfrac{1}{8}\cum{Y_{\lambda1}^{[3]};Y_{\lambda2}^{[2]};Y_{\lambda5}}
  +\sfrac{5}{54}\cum{Y_{\lambda1}^{[3]};Y_{\lambda3}^{[3]}}
  +\sfrac{5}{16}\cum{Y_{\lambda1}^{[2]};Y_{\lambda2}^{[3]};Y_{\lambda4}}
+\sfrac{1}{64}\cum{Y_{\lambda2}^{[6]}}
\nonumber\\*&\quad
  +\sfrac{5}{4}
  \cum{Y_{\lambda1}^{[2]};Y_{\lambda2}^{[2]};Y_{\lambda3}^{[2]}}
  +\sfrac{5}{16}
  \cum{Y_{\lambda1};Y_{\lambda2}^{[4]};Y_{\lambda3}}
  +\sfrac{5}{12}\cum{Y_{\lambda1}^{[3]};Y_{\lambda2};Y_{\lambda3};Y_{\lambda4}}.
\label{c6gen}
\end{align}
To evaluate these expressions, we need the explicit expressions for
the $Y_{\lambda\gamma}$. Since the $Y_{\lambda\gamma}$ are simply the
$\gamma$th derivative of $\Delta x_{\lambda1}$, they can be found by
straightforward differentiation (cf.\ Eqs.\,\eqref{eq9} and
\eqref{Deltaxdef}). The resulting expressions are polynomials in the
velocities of the particles\cite{VanZonCohen05}. Below, it will turn
out that only the highest power of the velocities in the expression of
each $Y_{\lambda\gamma}$ leads to a non-zero contribution to
$c^\lambda_4$ and $c^\lambda_6$. It suffices therefore to write only
the highest powers in the velocities for the $Y_{\lambda\gamma}$,
i.e.,
\begin{align} 
  Y_{\lambda1} &=  v_{\lambda1x}
\label{Y1}
\\
  Y_{\lambda2} &= -\frac{1}{m_\lambda}\dd{U}{x_{\lambda1}}
\label{Y2}
\\
  Y_{\lambda3} &= -\frac{1}{m_\lambda}
  \sum_{\mu,j}
  \dd{{}^2U}{x_{\lambda1}\partial\mathbf r_{\mu j}} \cdot \mathbf v_{\mu j}
\label{Y3}
\\
  Y_{\lambda4} &= 
  -\frac{1}{m_\lambda}\sum_{\mu,j}
  \sum_{\nu,k}
  \dd{{}^3U}{x_{\lambda1}\partial\mathbf r_{\mu j}\mathbf r_{\nu k}} :
  \mathbf v_{\mu j} \mathbf v_{\nu k}
  + \mathcal O(v^0)
\label{Y4}
\\
  Y_{\lambda5} &= 
  -\frac{1}{m_\lambda}
  \sum_{\mu,j}
  \sum_{\nu,k}
  \sum_{\kappa,\ell}
  \dd{{}^4U}{x_{\lambda1}\partial\mathbf r_{\mu j}\mathbf r_{\nu k}
    \mathbf r_{\kappa\ell}} :
  \mathbf v_{\mu j} \mathbf v_{\nu k} \mathbf v_{\kappa\ell}
  + \mathcal O(v^1)
\label{Y5}
\end{align}\begin{align}
  Y_{\lambda6} &= 
  -\frac{1}{m_\lambda}
  \sum_{\mu,j}
  \sum_{\nu,k}
  \sum_{\kappa\ell}
  \sum_{\rho n}
  \dd{{}^5U}{x_{\lambda1}\partial\mathbf r_{\mu j}\mathbf r_{\nu k}
  \mathbf r_{\kappa\ell}\mathbf r_{\rho n}} :
  \mathbf v_{\mu j} \mathbf v_{\nu k} \mathbf v_{\kappa\ell} \mathbf
  v_{\rho n}
  + \mathcal O(v^2)
\label{Y6}
\\
  Y_{\lambda7} &= 
  -\frac{1}{m_\lambda}
  \sum_{\mu,j}\sum_{\nu,k}\sum_{\kappa\ell}\sum_{\rho n}\sum_{\tau p}
  \dd{{}^6U}{x_{\lambda1}\partial\mathbf r_{\mu j}\mathbf r_{\nu k}
  \mathbf r_{\kappa\ell}\mathbf r_{\rho n}\mathbf r_{\tau p}} :
  \mathbf v_{\mu j} \mathbf v_{\nu k} \mathbf v_{\kappa\ell} \mathbf
  v_{\rho n}\mathbf v_{\tau p}
  + \mathcal O(v^3)
\label{Y7}
\end{align}
where each sum over two indices denotes a sum over the components $A$
and $B$ for the Greek index and a sum over the particles of that
component for the Latin index, while $\mathcal O(v^n)$ represents
terms which are a polynomial of order $n$ in the velocities.

\subsection{Simplifications for equilibrium systems}
\label{sec:eq}

In equilibrium, the velocities are independent Gaussian distributed
variables with zero mean (cf.\ Eqs.\,\eqref{rhoeq} and \eqref{Ham}), which
allows some simplifications in the expressions for $c^\lambda_4$ and
$c^\lambda_6$ in Eqs.\,\eqref{c4gen} and \eqref{c6gen}, respectively. These
simplification will not only lead to shorter expressions but will also
reduce the number of quantities inside each cumulant, i.e., it will
reduce the order of the cumulants. This is numerically advantageous
since higher order cumulants tend to require more statistics to keep
the error small.

The first simplification is that, given the Gaussian nature of the
velocities, Theorem A of Ref.\,\onlinecite{VanZonCohen05} can be applied to show
that the terms denoted by $\mathcal O(v^n)$ in
Eqs.\,\eqref{Y4}--\eqref{Y7} do not contribute to the right-hand side
of Eqs.\,\eqref{c4gen} and \eqref{c6gen}, because they contribute
cumulants which contain fewer powers of the velocity than the number
of velocity factors $Y_{\lambda1}=v_{x\lambda 1}$ in the cumulants,
and according to Theorem A, such cumulants are zero (see the Appendix
in Ref.\,\onlinecite{VanZonCohen05} for details).  On the other hand, the first
terms on the right-hand sides of Eqs.\,\eqref{Y4}--\eqref{Y7} contain
just enough powers of the velocities to match the number of factors of
$Y_{\lambda1}=v_{x\lambda 1}$ in the cumulants in Eqs.\,\eqref{c4gen}
and \eqref{c6gen} so that Theorem A does not apply and they might
yield a non-zero result.  Thus only these terms in
Eqs.\,\eqref{Y4}--\eqref{Y7} need to be taken into account.

The next simplification involves the average over the velocities,
which can be taken separately from the average over the positions
because of the factored form of the canonical equilibrium distribution
given in Eq.\,\eqref{rhoeq}. Thus, canonical averages can be taken in
two steps: first an average over velocities and then an average over
positions.  To apply this two-step process to cumulants, one needs to
relate the cumulants to averages. Using Eq.\,\eqref{cumintermsofmom},
the cumulants on the right-hand sides of Eqs.\,\eqref{c4gen} and
\eqref{c6gen} can be written in terms of moments which are simply
averages of products of factors of $Y_{\lambda\gamma}$.  For velocity
averages of products of independent Gaussian distributed velocities
with zero mean, we can use Wick's theorem which states that the
average can be obtained by pairing the velocities in all possible ways
and then taking the average for each pair separately. Note that the
average of two velocities $v_{\mu_1i_1}$ and $v_{\mu_2 i_2}$ is
\begin{equation}
  \av{v_{\mu_1i_1}v_{\mu_2 i_2}}_v
  =\frac{T}{m_{\mu_1}}\delta_{\mu_1\mu_2}\delta_{i_1i_2}.
\label{vv},
\end{equation}
where the subscript $v$ of the brackets indicates that only the
average over velocities is performed. Afterwards, the average over
positions, denoted by $\av{}_r$, still needs to be performed to
obtain the full average.

The straightforward method of writing the cumulants out in terms of
moments introduces a lot of subtractions terms, which can be largely
avoided by formulating a similar Wick's rule for cumulants. However,
the two-step nature of the averaging process, involving velocity as
well as position averages, is a complicating factor here.  Forgetting
for the moment about the position average, for Gaussian distributed
velocities, cumulants can be computed similarly as averages, i.e.\
using Eq.\,\eqref{vv}, with the distinction that there be only ``connected
contributions'', in the sense that the pairing of velocities be such
that all expressions in the cumulant are connected to each other.  To
give an example, for the cumulant $\cum{v_iv_j;v_kv_l}_v$, the
term $\textav{v_iv_j}_v\textav{v_kv_l}_v$ does not connect the
expressions $v_iv_j$ and $v_kv_l$, and therefore does not contribute,
while the terms $\textav{v_iv_k}_v\textav{v_jv_l}_v$ and
$\textav{v_iv_l}_v\textav{v_jv_k}_v$ do connect the two, so that
$\cum{v_iv_j;v_kv_l}_v=\textav{v_iv_k}_v\textav{v_jv_l}_v +
\textav{v_iv_l}_v\textav{v_jv_k}_v$.  However, when averaging with
$\rho_{eq}$ in Eq.\,\eqref{rhoeq}, there is a second, non-Gaussian, average,
namely, over the positions. As a consequence, although a term like
$\textav{\dd{{}^{2}U}{r_i\partial r_j}v_iv_j}_v
\textav{\dd{{}^{2}U}{r_k\partial r_l}v_kv_l}_v$ may seem
disconnected and therefore not to contribute to the cumulant
$\cum{\dd{{}^{2}U}{r_i\partial r_j}v_iv_j;
\dd{{}^{2}U}{r_k\partial r_l}v_kv_l}$, the second average over
positions will, as it were, reconnect the parts. One can show such
seemingly disconnected expressions (as far as the velocities are
concerned) still yield a contribution to the cumulant which is equal
to the position-cumulant of the factors, i.e.
$\cum{\textav{\dd{{}^{2}U}{r_i\partial r_j}v_iv_j}_v
;\textav{\dd{{}^{2}U}{r_k\partial
r_l}v_kv_l}_v}_r=\cum{\dd{{}^{2}U}{r_i\partial r_j}
;\dd{{}^{2}U}{r_k\partial
r_l}}_r\textav{v_iv_j}_v\textav{v_kv_l}_v$, where a subscript $r$
denotes a cumulant over the positions only.

With these rules on how to compute cumulants, we now return to the
expressions for $c^\lambda_4$ and $c^\lambda_6$ in Eqs.\,\eqref{c4gen}
and \eqref{c6gen}, respectively. One easily checks that to get
connected contributions, all the factors $Y_{\lambda1}=v_{x\lambda1}$
in the cumulants in Eqs.\,\eqref{c4gen} and \eqref{c6gen} must be
paired with velocities in the other $Y_{\lambda\gamma}$. If $n_1$ is
the number of factors of $Y_{\lambda1}$ in a cumulant, this introduces
a factor $n_1!$ due to the number of ways one can pair two sets of
$n_1$ velocities. Furthermore, because of the Kronecker delta's in
Eq.\,\eqref{vv}, all summations from Eqs.\,\eqref{Y3}--\eqref{Y7} can
easily be performed, and one finds
\begin{align} 
c^\lambda_4 &=\frac{1}{m_\lambda^4}\Bigg[
-\frac{T^3}{5}\rcum{\dd[4]{U}{x_{\lambda1}}}
+\frac{T^2}{3}\rcum{\bigg(\dd[2]{U}{x_{\lambda1}}\bigg)^{[2]}}
+\frac{T^2}{2}\rcum{\dd{U}{x_{\lambda1}};\dd[3]{U}{x_{\lambda1}}}
\nonumber\\*&\qquad
-\frac{T}{2}\rcum{\bigg(\dd{U}{x_{\lambda1}}\bigg)^{[2]};\dd[2]{U}{x_{\lambda1}}}
+\frac{1}{16}\rcum{\bigg(\dd{U}{x_{\lambda1}}\bigg)^{[4]}}
\Bigg]
\label{c4eq}
\end{align}
\begin{align}
   c^\lambda_6 &=
\frac{1}{m_\lambda^6}\Bigg[
  -\frac{T^5}{7}\rcum{\dd[6]{U}{x_{\lambda1}}}
  +\frac{T^4}{2}\rcum{\dd{U}{x_{\lambda1}};\dd[5]{U}{x_{\lambda1}}}
  +T^4\rcum{\dd[2]{U}{x_{\lambda1}};\dd[4]{U}{x_{\lambda1}}}
\nonumber\\* &\qquad
  +\frac{5T^4}{8}\rcum{\bigg(\dd[3]{U}{x_{\lambda1}}\bigg)^{[2]}}
  -\frac{3T^3}{4}\rcum{\bigg(\dd{U}{x_{\lambda1}}\bigg)^{[2]};\dd[4]{U}{x_{\lambda1}}}
  -\frac{5T^3}{9}\rcum{\bigg(\dd[2]{U}{x_{\lambda1}}\bigg)^{[3]}}
\nonumber\\* &\qquad
  +\frac{5T^2}{8}\rcum{\bigg(\dd{U}{x_{\lambda1}}\bigg)^{[3]};\dd[3]{U}{x_{\lambda1}}}
  -\frac{1}{64}\rcum{\bigg(\dd{U}{x_{\lambda1}}\bigg)^{[6]}}
\nonumber\\* &\qquad
  +\frac{5T^2}{2}\rcum{\bigg(\dd{U}{x_{\lambda1}}\bigg)^{[2]};\bigg(\dd[2]{U}{x_{\lambda1}}\bigg)^{[2]}}
  -\frac{5T}{16}
  \rcum{\bigg(\dd{U}{x_{\lambda1}}\bigg)^{[4]};\dd[2]{U}{x_{\lambda1}}}
\nonumber\\* &\qquad
  -\frac{5T^3}{2}\rcum{\dd{U}{x_{\lambda1}};\dd[2]{U}{x_{\lambda1}};\dd[3]{U}{x_{\lambda1}}}
\Bigg].
\label{c6eq}
\end{align}
Here the same notation has been used as explained below
Eq.\,\eqref{Clambdaj} and in \ref{A}.

The above expressions can still be further simplified for systems in
canonical equilibrium, using the following identity due to
Yvon\cite{yvon,vanzonschofield}
\begin{equation}
  \av{\dd{U}{x_{\lambda1}} B}_r = T \av{\dd{B}{x_{\lambda1}}}_r,
\label{Sears}
\end{equation}
for any function $B$ of the position of the particles, as can be
proved by partial integration. While we will not present the lengthy
details here, this identity can be used to find linear relations
between the expressions on the right-hand sides of Eqs.\,\eqref{c4eq2}
and \eqref{c6eq2}, which allow us to rewrite the expressions for
$c^\lambda_4$ and $c^\lambda_6$ in a variety of ways. Among those, we
choose
\begin{align}
  c^\lambda_4 &=\frac{1}{m_\lambda^4}\Bigg[
-\frac{T^3}{80}\rcum{\dd[4]{U}{x_{\lambda1}}}
+\frac{T^2}{48}\rcum{\left(\dd[2]{U}{x_{\lambda1}}\right)^{[2]}}
\Bigg]
\label{c4eq2}
\\
  c^\lambda_6 &=\frac{1}{m_\lambda^6}\Bigg[
  -\frac{T^5}{448}\rcum{\dd[6]{U}{x_{\lambda1}}}
  +\frac{T^4}{64}\rcum{\dd[2]{U}{x_{\lambda1}};\dd[4]{U}{x_{\lambda1}}}
  -\frac{5T^3}{576}\rcum{\bigg(\dd[2]{U}{x_{\lambda1}}\bigg)^{\![3]}}
\Bigg],
\label{c6eq2}
\end{align}
These equations require at most second and third order cumulants,
respectively, which is advantageous since numerically higher order
cumulants tend to produce larger statistical errors.  They agree with
the expressions found by Sears for a one-component fluid\cite{sears}.
Note that in the special case of a harmonic potential, derivatives
higher than the second vanish, so that then for $c^\lambda_4$ and
$c^\lambda_6$ only the last terms in Eqs.\,\eqref{c4eq2} and
\eqref{c6eq2}, respectively, remain, which only involve the cumulants
of the second derivative of the potential.  Since the second
derivative is constant for a harmonic potential, these cumulants are
zero as well, so that the coefficients $c^\lambda_4$ and $c^\lambda_6$
are zero, as expected for a linear system.

With this background, next, we will present the results of the
numerical evaluation of the coefficients $c_4^\lambda$ and
$c_6^\lambda$ for a number of equilibrium systems by means of
molecular dynamics simulations, in order to estimate the time scales
$\tau^\lambda_G$ and $\tau_*^\lambda$ which indicate where one could
suppose that the first term alone (i.e.\ the leading Gaussian) or the
first few terms (i.e.\ the Gaussian plus corrections) of the series in
Eq.\,\eqref{eq8} can be used as a good approximation to the full Van Hove
self-correlation function.

\section{Simulation results}

\label{results}

\subsection{Single component Lennard-Jones fluid}

\label{single}

\begin{figure}[t]
\centerline{
\includegraphics[angle=-90,width=0.5\textwidth]{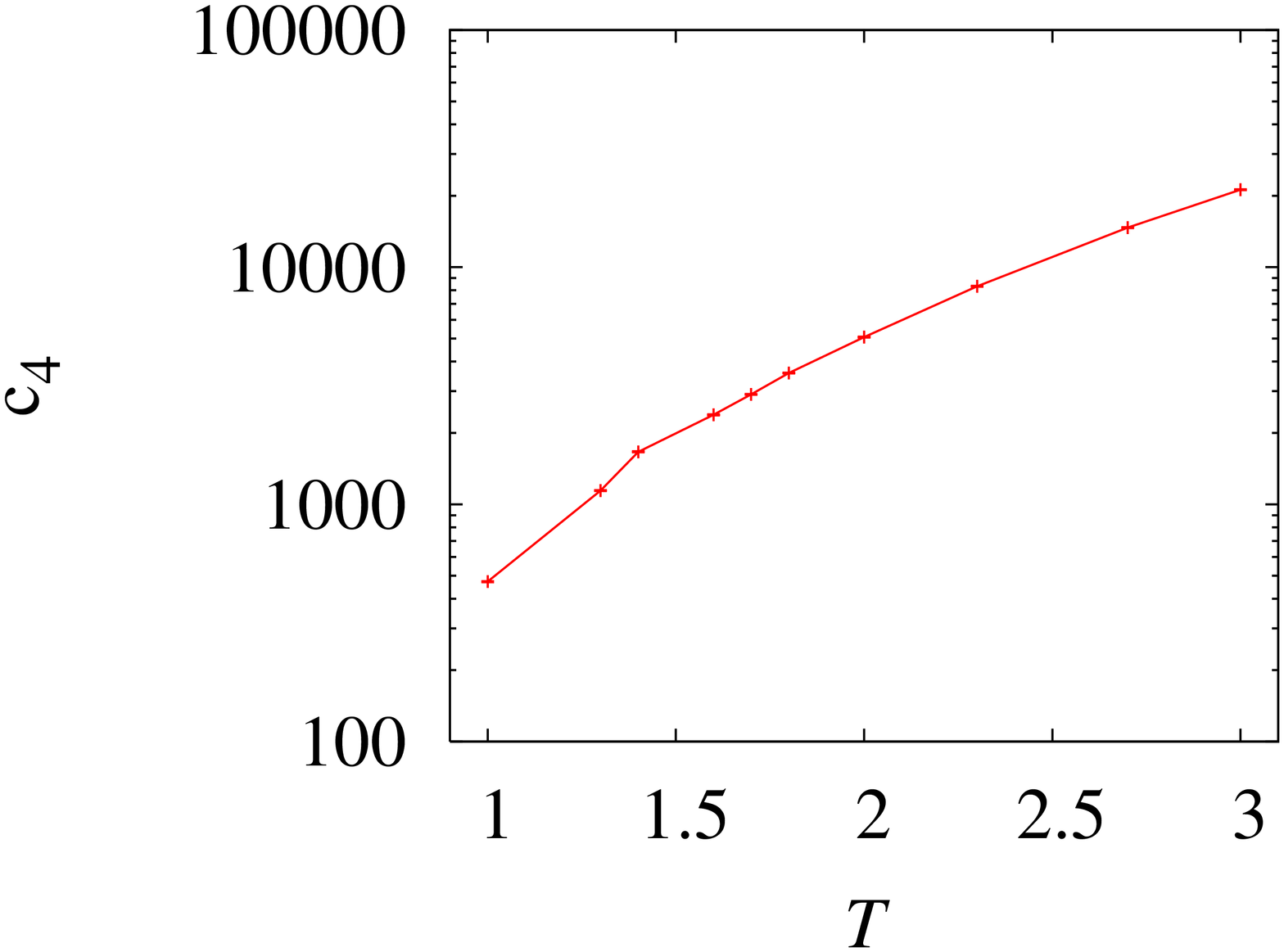}%
\includegraphics[angle=-90,width=0.5\textwidth]{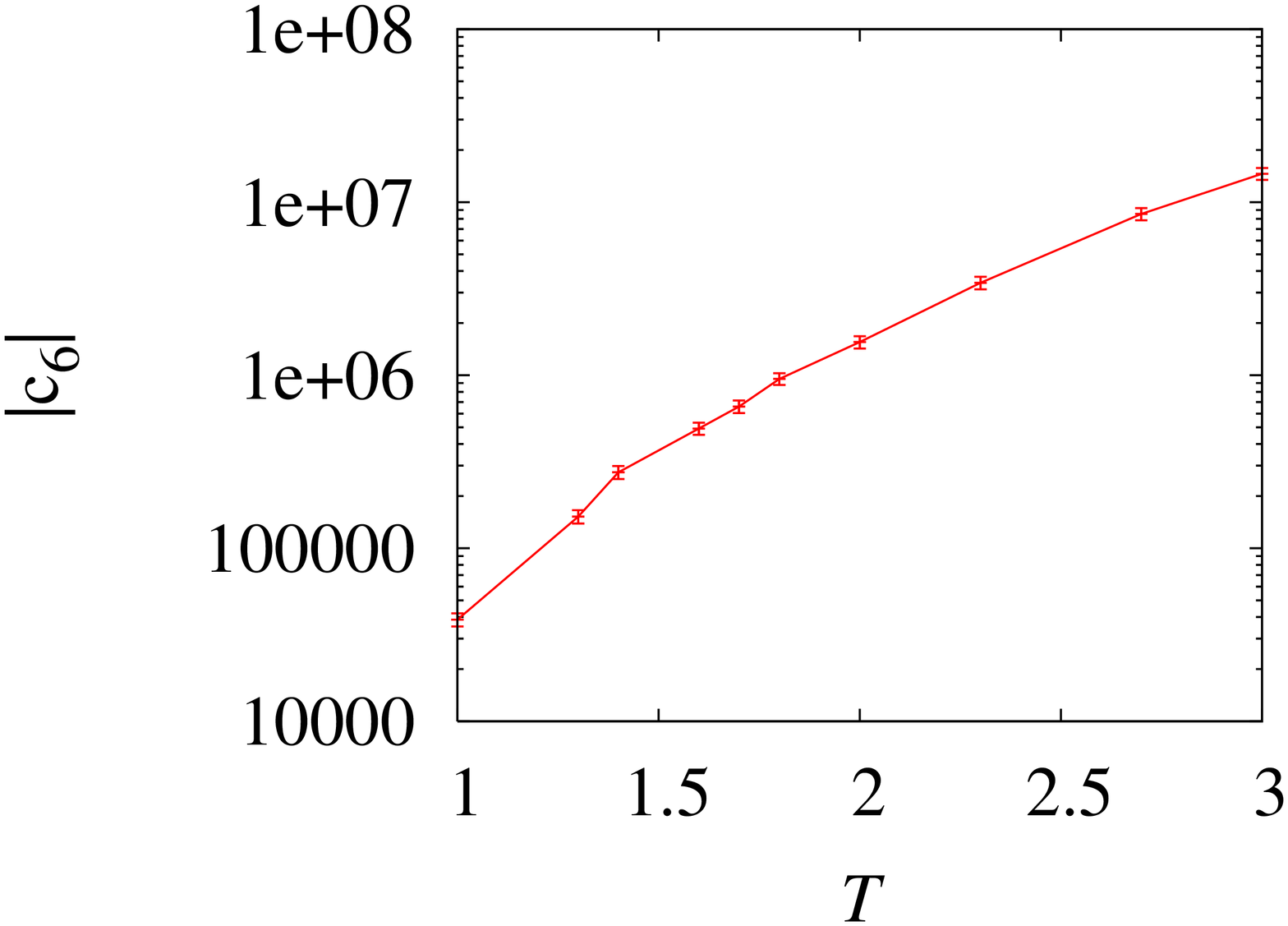}}
\caption{The coefficients $c_4$ (on the left) and $c_6$ (on the right)
as a function of temperature $T$ for an equilibrium single component
LJ fluid with density $\rho=0.8$.  These results are from a MD
simulation with $N=100$ particles, with periodic boundary
conditions. All quantities are in the LJ units defined in
Sec.\,\ref{systems}.}
\label{equilC4C6}
\end{figure}

In this section, we present the numerical result for $c_4$ and $c_6$
(cf.\ Eqs.\,\eqref{c4eq2} and \eqref{c6eq2}) and the resulting time
scales $\tau_G$ and $\tau_*$ (cf.\ Eqs.\,\eqref{tauG} and
\eqref{taustar}) for a single component fluid of $N=N_A$ LJ particles
with periodic boundary conditions in a box of linear size $L=5$ (in LJ
units). Note that we have omitted the component-superscript $\lambda$
here because there is only one component.  The results were obtained
from molecular dynamics (MD) simulations, for which the initial
conditions were drawn from the canonical distribution by employing an
isokinetic Gaussian thermostat\cite{EvansMorriss} during the
equilibration stage, while the runs themselves were done at constant
volume and energy.  In the simulation, a potential cutoff of
$r_c=2.5\sigma$ was used and the equations of motion were integrated
using the Verlet algorithm \cite{frenkel} with a time step of 2
femtoseconds.

\begin{figure}[t]
\centerline{\includegraphics[angle=-90,width=0.83\textwidth]{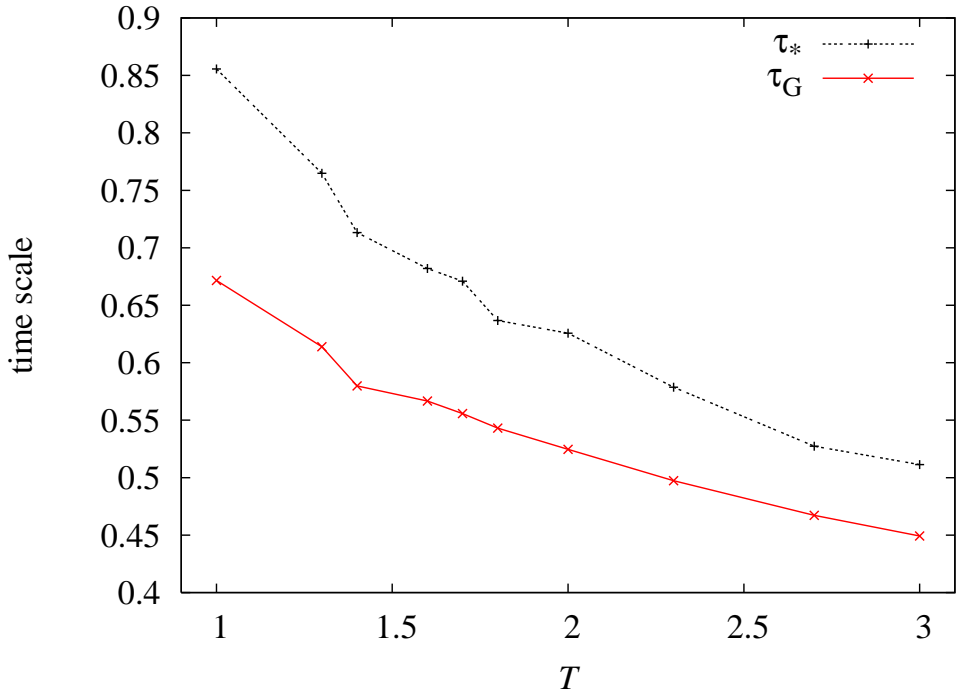}}
\caption{The critical time scales $\tau_G$ and $\tau_*$ at which the
series in Eq.\,\eqref{eq8} for the Van Hove self-correlation function of an
equilibrium single component fluid could be supposed to be practicable
(cf.\ Sec.\,\ref{timescales}, below Eqs.\,\eqref{tauG} and \eqref{taustar}) as a
function of temperature $T$ for a density $\rho=0.8$. Note that the
physical time scales in picoseconds can be calculated by multiplying
both $\tau_*$ and $\tau_G$ by the LJ unit time $\tau_{LJ}=2.16$ ps.}
\label{timescale}
\end{figure}
      
Since $\tau_G$ and $\tau_*$ will depend on temperature and density, it
is of interest to study the dependence of $c_4$ and $c_6$ as a
function of these two parameters. We studied the temperature
dependence by keeping $N$ and $\rho$ fixed to $100$ and $0.8$,
respectively, while temperature values ranging from 1 to 3 were used.
For each of these parameter values, data were accumulated once
equilibrium had been attained in the simulation and collected every 2
ps in a 8 ps long run, yielding five points per run. This was repeated
for 2000 different initial conditions (yielding 10,000 points per
temperature) for each temperature value and the results for $c_4$ and
$c_6$ were averaged over these 2000 runs. To decrease the statistical
errors even further, we averaged over all particles of the same kind
(i.e.\ replacing the index $1$ in Eqs.\,\eqref{c4eq2} and
\eqref{c6eq2} by any index $i$ and averaging the results) as well as
over the three directions of space (i.e. replacing $x$ by $y$ and $z$
in Eqs.\,\eqref{c4eq2} and \eqref{c6eq2} and averaging).

The resulting behaviour of $c_4$ and $c_6$ as a function of
temperature is shown in Fig.\,\ref{equilC4C6}.  The data for $c_4$ in the
left panel of Fig.\,\ref{equilC4C6} are consistent with the preliminary
data that were presented in Ref.\,\onlinecite{BrusselsPaper}.  Note that in
Fig.\,\ref{equilC4C6}, the absolute value of the coefficient $c_6$ has been
plotted. The reason is that the values of $c_6$ that are found in the
simulations are always negative. In Fig.\,\ref{timescale}, we plotted the
resulting time scales $\tau_G$ and $\tau_*$ (cf.\ Eqs.\,\eqref{tauG}
and \eqref{taustar}) as a function of temperature. We see that by
increasing the temperature, we moderately decrease these time scales
from roughly 2 ps to 1 ps, which are the estimates for the time scales
up to which the series in Eq.\,\eqref{eq8} could be supposed to give
an accurate approximation to $G^\lambda_s$.

\begin{figure}[t]
\centerline{\includegraphics[angle=-90,width=0.83\textwidth]{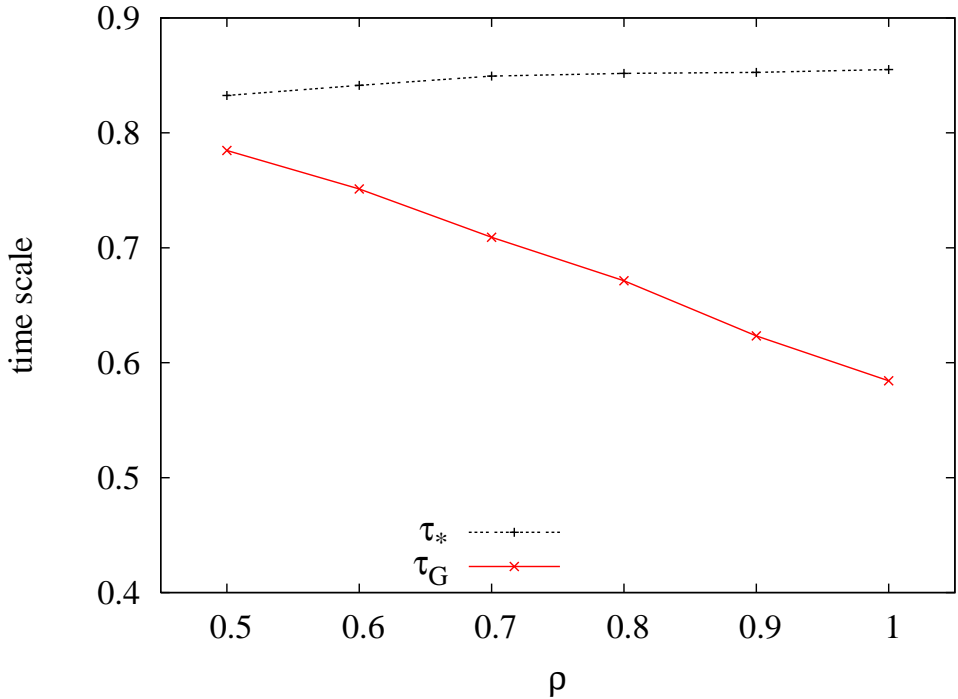}}
\caption{The critical time scales $\tau_G$ and $\tau_*$ at which the
  series in Eq.\,\eqref{eq8} for the equilibrium single component
  fluid could be supposed to be useful (cf.\ Sec.\,\ref{timescales}, below
  Eqs.\,\eqref{tauG} and \eqref{taustar}) as a function of the density
  $\rho$ for fixed temperature $T=1.0$. Note that the physical time
  scale in picoseconds can be calculated by multiplying both $\tau_*$
  and $\tau_G$ by the LJ unit time $\tau_{LJ}=2.16$ ps.}
\label{timescalerho}
\end{figure}

The density dependence of $c_4$ and $c_6$ was also investigated using
the same setup, but keeping the temperature fixed at $T=1.0$, while
the density ranged from $\rho=0.5$ to $\rho=1.0$. The resulting time
scales $\tau_G$ and $\tau_*$ as a function of density are plotted in
Fig.\,\ref{timescalerho}.  While both timescales remain on the order of one
or two picoseconds under changes of the density, we see that the two
time scales $\tau_G$ and $\tau_*$ behave quite differently; whereas
the time scale $\tau_G$ decreases moderately with increasing density,
indicating that the first correction term in Eq.\,\eqref{eq8} becomes
important somewhat sooner for higher than for lower densities, the
time scale $\tau_*$ is virtually constant as a function of density and
bigger than $\tau_G$, indicating that the second correction term in
Eq.\,\eqref{eq8} becomes important at a slightly larger time scale. However,
the order of magnitude of these two time scales is so similar (i.e.\
both of picosecond order) that such a distinction does not appear to
be significant.

\subsection{Isotopic Lennard-Jones Binary Mixture}

Our investigation into the cumulants originated in the study of mass
transport in binary isotopic mixtures at short time
scales\cite{VanzonCohenarchive}, and hence we are interested in the
time scales $\tau^\lambda_G$ and $\tau^\lambda_*$ in binary isotopic
mixtures as well.  From the expressions for the time scales in
Eqs.\,\eqref{tauG} and \eqref{taustar} as well as for the coefficients
$c^\lambda_4$ and $c^\lambda_6$ in Eqs.\,\eqref{c4eq} and \eqref{c6eq},
respectively, one can readily deduce that $c^\lambda_4\propto
m_\lambda^{-4}$ and $c^\lambda_6\propto m_\lambda^{-6}$. Using this in
Eqs.\,\eqref{tauG} and \eqref{taustar}, one sees that the time scales
$\tau^\lambda_G$ and $\tau^\lambda_*$ simply scale as the square root
of the mass. The remaining parts of the coefficients only involve the
potential, which in an isotopic mixture is the same as for a pure LJ
system. Therefore, no new simulations are needed for this case; the
time scales are those of the pure LJ system, multiplied by the square
root of the mass ratio of the components and the original LJ
particles, i.e.:
\begin{align}
  \tau_G^\lambda &= \tau_G \sqrt{\frac{m_\lambda}{m}}
\\
  \tau_*^\lambda &= \tau_* \sqrt{\frac{m_\lambda}{m}}
\end{align}
where $m$ is the mass of the particles in a single component LJ fluid.

Since in Nature, there are no isotopes with large mass ratios, we
conclude that for isotopic binary mixtures the time scales at which
the series in Eq.\,\eqref{eq8} can be supposed to be useful are the same as
those for a single LJ fluid, i.e., of the order of a picosecond.

\subsection{Nanofluids}

\label{nano}

A nanofluid is a binary mixture of LJ fluid particles ($A$ particles)
and nanoparticles ($B$ particles). For such a mixture, the time scales
$\tau^A_G$ and $\tau^A_*$ and $\tau^B_G$ and $\tau^B_*$ need not be
the same. They were here investigated using the same approach as
above, but there are additional numerical challenges.  First of all,
for large $B$ particles, the typical relaxation and correlation times
(say of the particle velocity) grow with increasing $R$ due to the
increased inertia of the $B$ particle.  As a result, it takes longer
to equilibrate such a system, and one obtains fewer independent data
points per time unit. Secondly, since the $B$ particle is already
quite large, to surround it with a liquid-like fluid of $A$ particles
requires a large number of $A$ particles. This increase of the number
of particles causes a substantial slow down of the simulations. To
keep down the number of $A$ particles, one takes as few $B$ particles
as possible. This contributes to a third difficulty, namely, that for
the $B$ particles, there are fewer particles to average over, leading
to poorer statistics.

Given these difficulties, fewer runs can be performed in a reasonable
time for these systems and as a result the error bars on the data for
the $B$ particles are substantially larger than those for the $A$
particles and of the LJ fluids of the previous sections. Nonetheless,
we have been able to extract estimates for the timescales at which the
series in Eq.\,\eqref{eq8} could be supposed to be useful also for these
systems.

\begin{table}[t]
\centering
\begin{tabular}{cc||c|c|c}
&& $R=2$ & $R=4$ & $R=6$
\\\hline
$N_B=1$&$c_4^A$& $283.6\pm0.4$  & $288.4\pm0.5$                & $296 \pm 0.7$
\\
       &$c_6^A$&$-24524\pm477$  & $-24865\pm554$               & $-25369\pm 690$
\\
       &$c_4^B$&$0.036\pm0.002$ & $(29.3\pm1.7)\times10^{-6}$  & $(0.40\pm0.03)\times10^{-6}$
\\
       &$c_6^B$&$-0.066\pm0.037$& $(-1.2\pm0.6)\times10^{-6}$  & $(-1.6\pm 0.9)\times10^{-9}$
\\\hline
$N_B=2$&$c_4^A$& $284.8\pm0.7$  & $293.5\pm0.9$                & $308\pm1$
\\
       &$c_6^A$&$-24777\pm788$  & $-25100\pm751$               & $-26052\pm1209$
\\
       &$c_4^B$&$0.036\pm0.002$ & $(29\pm2)\times10^{-6}$      & $(0.52\pm0.05)\times10^{-6}$
\\
       &$c_6^B$&$-0.079\pm0.058$& $(-1.0\pm0.6)\times10^{-6}$  & $(-2.0\pm1.5)\times10^{-9}$
\\\hline
$N_B=3$&$c_4^A$&$289.1 \pm 0.9$ & $300 \pm 1$                  & $314\pm2$ 
\\
       &$c_6^A$&$-25500\pm1039$ & $-25717 \pm 1170$            & $-26368 \pm 1257$
\\
       &$c_4^B$&$0.041\pm0.002$ & $(41\pm2)\times10^{-6}$      & $(0.78 \pm 0.06)\times10^{-6}$
\\
       &$c_6^B$& $-0.13\pm0.10$ & $(-2.2\pm1.2)\times10^{-6}$  & $(-3\pm 2)\times10^{-9}$
\end{tabular}
\caption{The coefficients $c^\lambda_4$ and $c^\lambda_6$ for the LJ particles ($A$) and
the nanoparticles ($B$) in the nanofluid of
Sec.\,\ref{nano} at $T=1$.\label{c4c6nano}}
\end{table}

For the simulations of the nanofluid, two temperature values were
taken: a low temperature $T=1$ (corresponding to 122 Kelvin for Argon)
and a high temperature $T=3$ (366 Kelvin, chosen to be closer to room
temperature).  The simulated system contained $N_B=1$, 2 or 3
nanoparticles of size $R=2$, 4 or 6 (i.e., all nine combinations were
studied). The linear box size was $L=30$ so that the number density of
the nanoparticles had the values $\rho_B=3.7\times10^{-5}$,
$7.4\times10^{-5}$ and $1.1\times10^{-4}$ for $N_B=1$, $2$ and $3$,
respectively.  To keep the properties of the LJ fluid in which the
nanoparticles are suspended constant, the remainder of the box was
filled with LJ particles with a fixed number density
$\rho_A=N_A/(L^3-\frac43\pi R^3N_B)$, which was, somewhat arbitrarily,
chosen to be 0.49, i.e. $N_A$ was chosen such that for given $L$, $R$
and $N_B$, $\rho_A$ was as close to 0.49 as possible.  This required
between $N_A=11,912$ and $N_A=13,227$ fluid LJ particles, depending on
$R$ and $N_B$. Note that even though the number densities of the
nanoparticles are small, by assigning a volume $\frac43\pi R^3$ to
each nanoparticle, one sees that the volume fraction ranges from
$0.124\%$ to $10\%$.  This is a realistic range, as experimental
volume fractions are of the order of $1\%$\cite{Choietal02}. We did
not investigate much higher volume fractions to avoid possible
complicating effects such as aggregation of the nanoparticles.

\begin{table}[t]
\centering
\begin{tabular}{cc||c|c|c}
       &       & $R=2$                   & $R=4$                     & $R=6$
\\\hline
$N_B=1$&$c_4^A$&  $9042\pm 7$            & $9034\pm7$                & $9123 \pm 6$
\\
       &$c_6^A$&$(-6.4\pm0.12)\times10^6$&$(-6.3\pm0.11)\times10^{6}$&$(-6.4\pm0.11)\times10^6$
\\
       &$c_4^B$&  $17\pm1$               & $(22\pm2)\times10^{-3}$   & $(357\pm54)\times10^{-6}$
\\
       &$c_6^B$&$-442\pm291$             & $(-11\pm17)\times10^{-3}$ & $(-18\pm 76)\times10^{-6}$
\\\hline
$N_B=2$&$c_4^A$& $9049\pm8$              & $9163\pm8$                & $9296\pm8$
\\
       &$c_6^A$&$(-6.4\pm0.13)\times10^6$& $(-6.5\pm0.14)\times10^6$ & $(-6.6\pm0.12)\times10^6$
\\
       &$c_4^B$& $17\pm1$                & $(23\pm2)\times10^{-3}$   & $(390\pm43)\times10^{-6}$
\\
       &$c_6^B$&$-434\pm192$             & $(-13\pm17)\times10^{-3}$ & $(-18\pm62)\times10^{-6}$
\\\hline
$N_B=3$&$c_4^A$&$9087 \pm 8$             & $9203 \pm 6$              & $9455\pm160$ 
\\
       &$c_6^A$&$(-6.4\pm0.12)\times10^6$& $(-6.5\pm0.1)\times10^6$  & $(-6.8\pm0.7)\times10^6$
\\
       &$c_4^B$&$17\pm1$                 & $(22\pm1)\times10^{-3}$   & $(468\pm110)\times10^{-6}$
\\
       &$c_6^B$& $-425\pm198$            & $(-11\pm11)\times10^{-3}$ & $(-4.3\pm105)\times10^{-6}$
\end{tabular}
\caption{The coefficients $c^\lambda_4$ and $c^\lambda_6$ for the LJ particles ($A$) and the
nanoparticles ($B$) in the nanofluid of Sec.\,\ref{nano} at $T=3$.}
\label{c4c6nanoT3}
\end{table}

For the systems with $1$ nanoparticle, 100 runs were performed for
each of the two temperature values $T=1$ and $T=3$, where first the
system was equilibrated using an isokinetic Gaussian thermostat, and
then the system was run for 8 ps during which the quantities appearing
in Eqs.\,\eqref{c4eq2} and \eqref{c6eq2} were measured. For the systems with
$N_B=2$, 50 runs were performed and for those with $N_B=3$ the number
of runs was 34 (for each temperature value).  Because of the
isokinetic Gaussian thermostat, the average over these runs
approximates the average over the canonical distribution in
Eq.\,\eqref{rhoeq}.
 
The resulting values for $c^\lambda_4$ and $c^\lambda_6$ are shown in
Tables \ref{c4c6nano} and \ref{c4c6nanoT3} for $T=1$ and $T=3$,
respectively. From $c^\lambda_4$ and Eq.\,\eqref{tauG} we find the
timescales $\tau^\lambda_G$, which are listed in Tables \ref{ttau} and
\ref{ttauT3} for $T=1$ and $T=3$, respectively.  In Tables
\ref{c4c6nano} and \ref{c4c6nanoT3}, one notices the large error
estimates for $c^B_6$ (whose values are negative as in the pure LJ
case), which may seem to make it hard to draw conclusions from those
data. However, according to Eq.\,\eqref{c6eq2} we only need the square
root of this number to estimate $\tau^B_*$, leading to a reduction of
the relative error by one half, which explains why the results for
$\tau^B_*$ given in Tables \ref{ttau} and \ref{ttauT3} are still
reasonable \emph{order of magnitude} estimates for all cases except
for the combination of physical parameters $R=6$ and $T=3$.

We see from Tables \ref{ttau} and \ref{ttauT3} that for the LJ fluid
particles ($A$) surrounding the nanoparticles, both time scales
$\tau_G^A$ and $\tau_*^A$ (below which which the expansion of the Van
Hove self-correlation function around a Gaussian as in
Eq.\,\eqref{eq8} may be useful) are on the order of one or two
picoseconds. While they decrease moderately with increasing
temperatures, these time scales are relatively insensitive both to the
radius and to the density of the nanoparticles, and are in fact close
to their values in the absence of nanoparticles
(cf. Fig.\,\ref{timescale}), which were also on the order of one to two
picoseconds.

In contrast to this, Tables \ref{ttau} and \ref{ttauT3} shows that the
time scales below which the expansion of the Van Hove self-correlation
function of the nanoparticles ($B$) around a Gaussian could be
supposed to be practicable, is considerably larger than for the fluid
particles, and, in fact, increases with the radius of the
nanoparticles up to as much as a factor five for $T=3$ and a factor
ten for $T=1$ for the largest nanoparticle size studied. The
timescales decrease upon increasing the density of the nanoparticles,
but by a lesser amount, so that the overall timescale below which
Eq.\,\eqref{eq8} is useful is still on the order of five picoseconds
for $T=3$ and on the order of ten picoseconds for $T=1$.

\begin{table}[t]
\centering
\begin{tabular}{cc||c|c|c}
&& $R=2$ & $R=4$ & $R=6$
\\\hline
$N_B=1$&$\tau_G^A$&$0.763\pm0.001$ & $0.759\pm0.001$ & $0.755\pm0.001$
\\
       &$\tau_*^A$&$0.833\pm0.008$ & $0.834\pm0.009$ & $0.837\pm0.011$
\\
       &$\tau_G^B$& $2.40\pm0.03$  & $5.28\pm 0.08$  & $8.45\pm0.16$
\\
       &$\tau_*^B$& $1.9\pm0.5$    & $4.7\pm1.2$     & $8\pm 2$
\\\hline
$N_B=2$&$\tau_G^A$&$0.762\pm0.001$ & $0.756\pm0.001$ & $0.747\pm0.001$
\\
       &$\tau_*^A$&$0.830\pm0.013$ & $0.838\pm0.013$ & $0.842\pm0.019$
\\
       &$\tau_G^B$&$2.40\pm0.03$   & $5.29\pm0.09$   & $7.91\pm0.19$
\\
       &$\tau_*^B$& $1.9\pm0.8$    & $5.2\pm1.6$     & $8\pm 3$
\\\hline
$N_B=3$&$\tau_G^A$&$0.759\pm0.001$ & $0.752\pm0.001$ & $0.744\pm0.001$
\\
       &$\tau_*^A$&$0.825\pm0.017$ & $0.837\pm0.019$ & $0.84\pm0.02$
\\
       &$\tau_G^B$& $2.32\pm0.03$  & $4.85\pm0.06$   & $7.15\pm0.14$
\\
       &$\tau_*^B$& $1.5\pm0.6$    & $4.1\pm1.1$     & $8\pm3$
\end{tabular}
\caption{The time scales $\tau^\lambda_G$ and $\tau^\lambda_*$ (in LJ
  units) as follow from $c^\lambda_4$ and $c^\lambda_6$ according to
  Eqs.\,\eqref{tauG} and \eqref{taustar} for the LJ particles ($A$) and the
  nanoparticles ($B$) in the nanofluid of Sec.\,\ref{nano} at $T=1$,
  respectively. Note that the physical time scale in picoseconds can
  be calculated by multiplying $\tau_*$ and $\tau_G$ by the LJ unit
  time $\tau_{LJ}=2.16$ ps.}
\label{ttau}
\end{table}

\section{Conclusions}

\label{conclusions}

We have investigated the short time behaviour of the Van Hove
self-correlation function. According to Eq.\,\eqref{eq8}, for short times,
the Hove self-correlation function can be expressed as a Gaussian plus
corrections, which are proportional to increasing powers of $t$.  For
short times, this can be re-expressed by the series in Eq.\,\eqref{eq8},
which is useful provided the contributions of the correction terms are
small. From the form of these correction terms in Eq.\,\eqref{eq8}, one sees
that they are small at time scales smaller than some critical time
scale ($\tau_G$). In this paper, this time scale was investigated for
a number of LJ and LJ-based systems.  We found that a decrease of the
magnitude of the terms in the series Eq.\,\eqref{eq8} occurs below and up to
the picosecond time scales for LJ fluid particles and up to the ten
picosecond time scale for nanoparticles.

Two time scales were in fact calculated: one, denoted by $\tau_G$,
estimates when the first correction term to the Gaussian distribution
will be small, and the other, denoted by $\tau_*$, estimates the time
at which the second correction term is as big as the first one.  The
larger these time scales, the better, since this means that the
expansion in Eq.\,\eqref{eq8}, i.e., the Gaussian plus two correction terms,
or perhaps even just the simple Gaussian prefactor, can be used for
all time scales below (and possibly up to) $\tau_G$ and $\tau_*$. Note
that if these time scales are of similar order of magnitude, as they
turned out to be, then they could also be viewed as a possible
estimate of the radius of convergence of the series in Eq.\,\eqref{eq8}.

We first investigated the coefficients for the equilibrium pure LJ
fluid as a function of temperature and concluded that both time scales
$\tau_G$ and $\tau_*$ are reduced as a function of increasing
temperature from about 2 picoseconds to 1 picosecond. As a function of
density, our two estimates of the time scales behave
differently. While $\tau_G$ decreases by moderate amounts with
increasing density, $\tau_*$ stays roughly the same. In all cases
though, the timescales are of the order of a picosecond or more. One
can qualitatively understand the decreasing trend of the `Gaussian'
time scale $\tau_G$ for increasing densities, by realizing that the
forces between the particles perturb the short time ballistic motion
away from its Gaussian character. Since the forces are stronger at
higher densities, the deviations from Gaussian behaviour will then
occur earlier.

\begin{table}[t]
\centering
\begin{tabular}{cc||c|c|c}
&& $R=2$ & $R=4$ & $R=6$
\\\hline
$N_B=1$&$\tau_G^A$&$0.5560\pm0.0001$&$0.5561\pm0.0001$& $0.5547\pm0.0001$
\\
       &$\tau_*^A$&$0.503\pm0.005$  & $0.508\pm0.004$ & $0.508\pm0.004$
\\
       &$\tau_G^B$& $0.89\pm0.01$   & $1.75\pm 0.04$  & $2.7\pm0.1$
\\
       &$\tau_*^B$& $0.88\pm0.29$   & $2.4\pm1.9$     & $4\pm 8$
\\\hline
$N_B=2$&$\tau_G^A$&$0.5559\pm0.0001$&$0.5541\pm0.0001$& $0.5547\pm0.0001$
\\
       &$\tau_*^A$&$0.505\pm0.005$  & $0.504\pm0.005$ & $0.508\pm0.004$
\\
       &$\tau_G^B$&$0.89\pm0.01$    & $1.73\pm0.03$   & $2.62\pm0.07$
\\
       &$\tau_*^B$& $0.88\pm0.20$   & $2.2\pm1.5$     & $4\pm 7$
\\\hline
$N_B=3$&$\tau_G^A$&$0.5553\pm0.0001$&$0.5535\pm0.0001$& $0.5498\pm0.0001$
\\
       &$\tau_*^A$&$0.507\pm0.005$  & $0.504\pm0.004$ & $0.50\pm0.25$
\\
       &$\tau_G^B$& $0.89\pm0.01$   & $1.74\pm0.02$   & $2.5\pm0.15$
\\
       &$\tau_*^B$& $0.90\pm0.21$   & $2.4\pm1.2$     & $9\pm114$
\end{tabular}
\caption{The time scales $\tau^\lambda_G$ and $\tau^\lambda_*$ (in LJ
  units) as follow from $c^\lambda_4$ and $c^\lambda_6$ according to
  Eqs.\,\eqref{tauG} and \eqref{taustar} for the LJ particles ($A$)
  and the nanoparticles ($B$), respectively, in the nanofluid of
  Sec.\,\ref{nano} at a temperature of $T=3$. Note that the physical time
  scale in picoseconds can be calculated by multiplying $\tau_*$ and
  $\tau_G$ by the LJ unit time $\tau_{LJ}=2.16$ ps.}
\label{ttauT3}
\end{table}

In mixtures, there is a Van Hove self-correlation function for each
component, and correspondingly, the time scales depend on the
component whose Van Hove self-correlation function is studied, which
is represented by a superscript $\lambda=A$ or $B$ on $\tau_G$ and
$\tau_*$.  We deduced for a binary isotopic mixture that the time
scales $\tau_G^\lambda$ and $\tau_*^\lambda$ on which Eq.\,\eqref{eq8}
could be supposed to be useful, simply scale as the square root of the
mass $m_\lambda$ of the component $\lambda$.  As said before, since in Nature,
isotopes do not have very large mass ratios, for isotopic binary
mixtures the time scales at which the series in Eq.\,\eqref{eq8} is
useful are of the same order of magnitude as for a one-component
fluid, i.e., of the order of a picosecond.

Finally, we studied these time scales in a recently proposed model of
a nanofluid\cite{VanZonunpublished}, and found that the time scales
are there of the order of five to ten picoseconds for the
nanoparticles (decreasing with temperature and increasing with
radius), while for the fluid particles in that model the time scale is
still on the order of a picosecond. The difference in time scales
could be due to the larger mass of the nanoparticles, causing the
forces to have less influence on their velocities, which therefore
remain close to their original (Gaussian) distribution for a longer
time than in a LJ fluid. It is then no surprise that the distribution
of displacements for nanoparticles can be described by a Gaussian at
longer time scales than for the lighter fluid particles.

One may wonder whether the time scales found in this paper are not so
short that the classical description on which they were based breaks
down.  A simple estimate of the time scale at which appreciable
quantum effects can be expected is given by $\hbar/k_BT$, where
$\hbar$ is Planck's constant divided by $2\pi$. At room temperature,
this is equal to about 25 femtoseconds. Note that all of the time
scales found in this paper were at picosecond or at tens of picosecond
scales, i.e., well above this quantum time scale.

Although our results for the time scales $\tau_G^\lambda$ and
$\tau_*^\lambda$ are only estimates, they are encouraging for the
possible application of a Green's function approach to small scale
nanometre length and picosecond time scales, since the Van Hove
self-correlation functions are equilibrium versions of Green's
functions\cite{kincaid,kincHydro,kincNano,kincHeat,
VanzonCohenarchive}. Furthermore, it is expected that the time scales
for nonequilibrium systems are similar to those of equilibrium
systems, which were on the order of picoseconds for fluid particles
and on the order of ten picoseconds for nanoparticles. This suggests
that expansions of the form in Eq.\,\eqref{eq8} can be useful for the
Green's function approach for transport problems taking place at and
below picosecond time scales and at nanometre length scales in
equilibrium and near-equilibrium systems.

\acknowledgments

This work was supported by the Office of Basic Energy Sciences of the
US Department of Energy under grant number DE-FG-02-88-ER13847 and
under grant PHY-501315 of the Mathematical Physics program of the
National Science Foundation.

\appendix

\section{Moments and cumulants}

\label{A}

In this appendix we will briefly recall the definitions of the moments
and cumulants, and how they are related. For more details, see
Ref.\,\onlinecite{kampen}.

We first remark that multivariate moments and cumulants are simply
moments and cumulants of more than one variable. In general,
(multivariate) moments can be defined as follows. For a single random
variable $x$ with a distribution $f_1(x)$, the $n$th moment is
$\mu_n=\av{x^n}=\int\mathrm dx\:x^n f_1(x)$, while for a pair of
random variables $x_1$ and $x_2$ with a joint distribution
$f_2(x_1,x_2)$, the bivariate moments are
$\av{x_1^{n_1}x_2^{n_2}}=\int\mathrm dx_1\int\mathrm
dx_2\,x_1^{n_1}x_2^{n_2} f_2(x_1,x_2)$, and so on for multivariate
moments $\textav{x_1^{n_1}\cdots x_q^{n_q}}=\int\mathrm
dx_1\cdots\int\mathrm dx_q\,x_1^{n_1}\cdots x_q^{n_q}
f_q(x_1,\ldots,x_q)$. One defines the order of a multivariate moment
as the sum $\sum_{r=1}^q n_r$. For near-Gaussian (multivariate)
distributions, the cumulants are a more convenient way to characterize
the distribution than the moments, because the cumulants of order
higher than two are zero for a pure Gaussian. For a single variable
the general expression for the $n$th cumulant $\kappa_n$ in terms of
moments $\mu_{k\leq n}$ is
\begin{equation}
 \kappa_n = -n!
  \mathop{\sum_{\{p_\ell\geq 0\}}}_{\sum_{\ell=1}^\infty \ell p_\ell = n}
  \Big(\sum_{\ell=1}^\infty p_\ell-1\Big)!
  \prod_{\ell=1}^\infty \frac{\big[{-\mu_\ell}/{\ell!}\big]^{p_\ell}}{p_\ell!}.
\label{kappagen}
\end{equation}
In analogy with the notation $\mu_n=\textav{x^n}$ for moments of a
random variable $x$, one often uses the notation $\kappa_n=\cum{x^n}$
for its cumulants\cite{kampen}. Here, the superscript $n$ inside
the double brackets is not a power, as the example
$\cum{x^2}=\textav{x^2}-\textav{x}^2$ shows. To avoid confusion, we
denote instead the cumulants as $\cum{x^{[n]}}$.  Therefore, instead
of Eq.\,\eqref{kappagen} we can write
\begin{equation}
\cum{x^{[n]}} = -n!
\mathop{
\sum_{\{p_\ell\geq 0\}}
}_{\sum_{\ell=1}^\infty \ell p_\ell = n}
\Big(\sum_{\ell=1}^\infty p_\ell-1\Big)!
\prod_{\ell=1}^\infty \frac{\big[{-\textav{x^\ell}}/{\ell!}\big]^{p_\ell}
}{p_\ell!}.
\label{genkappa2}
\end{equation}
One can interpret the superscript $n$ between square brackets in this
expression as the number of `repetitions' of $x$.  Then, as an
alternative to Eq.\,\eqref{genkappa2}, one can define the cumulants
recursively as the average of the product of these repetitions minus
the product of lower order cumulants of all possible groupings of the
$n$ repetitions. For instance, for the third order cumulant of the
displacement one can write $\cum{x^{[3]}}= \textav{x^3} -
3\cum{x}\cum{x^{[2]}} - \cum{x}^3$, where the factor
three arises from the three ways in which one can group three
repetitions into a pair and a single repetition. This expression
contains the second order cumulant $\cum{x^{[2]}}$, which can be
written as $\cum{x^{[2]}}=\textav{x^2}-\cum{x}^2$,
while finally $\cum{x}=\textav{x}$, leading to
$\cum{x^{[3]}_{\lambda1}(t)}= \textav{x^3} -
3\textav{x}\textav{x^2} +2 \textav{x}^3$. This is a special case
of the general formula \eqref{genkappa2}.

Similarly to this univariate case, multivariate cumulants can be
represented in terms of the averages, in the following way
\cite{VanZonCohen05}:
\begin{align} 
\cum{x_1^{[n_1]};\dots;x_q^{[n_q]}}
 &= - n_1!\dots n_q!\!  
\mathop{
\sum_{\{p_{\{\ell\}}\geq 0\}}
}_{\sum_{\{\ell\}} \ell_{j} p_{\{\ell\}} =n_{j}}
\Big(\sum_{\{\ell\}} p_{\{\ell\}}-1\Big)!  \prod_{\{\ell\}}
\frac{1}{p_{\{\ell\}}!}
\bigg(-\frac{\textav{x_{1}^{\ell_1}\dots x_{{q}}^{\ell_{q}}}}{\ell_1!\dots\ell_{q}!}\bigg)^{p_{\{\ell\}}}\!\!.
\nonumber\\* &
\label{cumintermsofmom}
\end{align}
In this notation for the cumulants, quantities separated by semicolons
are treated as separate random variables and, as above, if a quantity
has a superscript within square brackets, it denotes that particular
number of repetitions of the quantity.  Some examples of multi-variate
cumulants in terms of multi-variate moments are
\begin{align} 
  \cum{x_1} &= \av{x_1}
\\
  \cum{x_1;x_2} &= \av{x_1x_2}-\av{x_1}\av{x_2}
\\
  \cum{x_1;x_2;x_3} &= \av{x_1x_2x_3}-\av{x_1x_2}\av{x_3}
                       -\av{x_1x_3}\av{x_2}-\av{x_1}\av{x_2x_3}
                       +2\av{x_1}\av{x_2}\av{x_3}
\end{align}

In the main text, the moments $\mu$ and cumulants $\kappa$ occurs as
moments and cumulants of the displacements of a single particles of a
specific component $\lambda$ in a time $t$, and therefore appear with
a superscript $\lambda$ (and an implicit time argument
$t$). Furthermore, multi-variate cumulants appear where the $x_\gamma$
are replaced by $Y_{\lambda\gamma}$, or by derivatives of the
potential, i.e.\ $\dd[\gamma]{U}{x_{\lambda 1}}$.

\end{document}